\documentclass[ aps, reprint, amsmath, amssymb ]{revtex4-1}
\usepackage{graphicx}
\usepackage{dcolumn}
\usepackage{bm}

\newcommand{\btt}{\mbox{\boldmath $\beta$}}

\begin {document}
\title{ On the quantum interpretation of the classical Schott term in the theory of radiation damping }
\author{Khokonov M.Kh.}
 \altaffiliation{Kabardino-Balkarian State University, Nalchik, Russian Federation.}
 \email{Electronic address: khokon6@mail.ru}     
\date{\today}

\begin{abstract}
The quantum interpretation of classical radiation reaction coming from the near electromagnetic self-force (the so-called ``Schott term'') is given for the first time. The analysis is based on the Landau--Lifshitz equation for classical ultrarelativistic motion including radiation reaction effects for the case of channeling radiation in oriented crystals in the wide region of electron energies from few MeV and up to hundreds GeV. This type of radiation is unique in that sense that it has a pronounced quantum character in two extreme cases of low and high energies. We show that quantum transitions between the transverse energy states of channeling particles represent the quantum analog of the classical Schott term. The impetus for this work was the recent reports on the feasibility of detecting experimentally the effects of the action of classical radiative self-force on the motion of electrons (positrons) channelled in oriented crystals at the CERN Secondary Beam Areas (SBA) beamlines, although the basic results of this work are valid for arbitrary ultrarelativistic motion.
\end{abstract}

\maketitle

\section{Introduction}

It has recently been shown that  classical radiation reaction effects, predicted by the Landau--Lifshitz equation (LL) \cite{Landau}, 
 can be measured  using presently available facilities at the CERN Secondary Beam Areas (SBA) beamlines 
 \cite{Ulrik_NC_2018}, \cite{Ulrik2017}. The authors of these papers adopted perspectives and presently feasible experimental setup to measure classical radiation reaction (RR) effects on the photon emission, generated in the interaction of ultrarelativistic electrons with an oriented crystal (the so-called channeling radiation).

Crystallographic axes and planes  provide electrostatic fields strong enough ($\sim 10^{11} \div 10^{12}$ V/cm) to make the RR effects visible. Channeling radiation turns out to be  promising to study the correspondence between the classical and quantum theories because quantum effects appear in the limits of relatively low electron energises of few MeV and at high energies above 100 GeV with pure classical nature of motion and radiation in between \cite{Beloshitsky1982}, \cite{Ulrik2005}. In the present letter this correspondence is analysed. 

All results of this work are valid for arbitrary ultrarelativistic motion but, as is mentioned above, we consider in detail the particular case of channeling radiation because it has a pronounced quantum character with individual lines in the emission spectrum at relatively low electron energies, and because channeling radiation can be handled theoretically in a comparatively simple way. In what follows we shall talk about electrons, while all conclusions are valid for positrons as well.

When an electron beam is incident on the crystal nearly parallel to an atomic axis (or plane), the successive correlated small-angle deflections that determine the transverse trajectory of the particle become important. In order to describe this type of scattering, Linhard introduced a continuous potential for an atomic axis (axial channeling) or atomic plane (planar channeling) \cite{Lindhard1965}. Such continuous potential depends only on the coordinates transverse to the axis (plane). As a result, the transverse energy and angular momentum with respect to the axis (for axial electron channeling) are preserved being integrals of the transverse motion.  The continuum potential approximation is valid when the initial angle of particle incidence with respect to the atomic string (plane) is smaller or of the same order of magnitude than some critical channeling angle $\theta_L = (2U_0/E)^{1/2}$, where $U_0$  is a continuous potential depth and $E$ is the electron energy. For axial  channeling $U_0 \simeq 2Ze^2/d \sim 10^2$ eV, $Z$ is the atomic number of crystal atoms, $d$ is a distance between atoms in the atomic string, $e$ is the electron charge. In the planar case   $U_0 \simeq$ 20-30 eV. 
The influence of radiation on the motion of multi - GeV electrons and positrons incident on single oriented crystals was studied both experimentally and theoretically about two decades ago, \cite{mkh1992}, \cite{EUggerhoj1997}, \cite{EUggerhoj2001}, under conditions when this influence is completely due to the hard photon emission corresponding to the irreversible energy loss term in classical radiation self-force expression (the so-called ``Li\'enard term'').  In the present Letter we shall mainly concentrate our attention on the opposite limit of relatively low energies when the influence of the reversible contribution to the radiative self-force is dominating.

The secondary effects of incoherent multiple scattering by thermal vibrations of crystal atoms and radiation damping lead to the change in transverse energy.  To study the effects of RR on electron motion we will take this change in transverse energy and angular momentum due to radiation as the basis of consideration. Multiple scattering will not be considered here although for realistic calculations this factor should be taken into account. 

Electromagnetic waves produced by powerful lasers can result in effect of intense radiation coming from relativistic electrons moving in the field of this waves \cite{Ritus1979},  \cite{mkh2002}, \cite{Keitel_RMP2012}. This provides perspectives for  possible  measurements of  laser-based classical RR  \cite{Neitz2014}, \cite{Ilderton_PRE2015}. In addition to the analysis of such perspectives  in comparison with oriented crystals given in \cite{Ulrik2017},  we emphasize that, in contrast with channeling, formation of the observable discrete quantum states in the transverse phase space of the electron moving in the laser field is not possible in principle. From this point of view the crystal-based tests of the so-called ``Schott term'' in the classical radiation reaction theory turn out to be more promising. On the other hand, the quantum strong field effects (like recoil and spin-flip) for peta-watt lasers can take place already at electron energies of few GeV, whereas in oriented crystals the energies above 100 GeV are required.

\section{The Landau-Lifshits equation in the static electric field}

The LL equation for electron moving in a static electric field  ${\bf E}$  is  (Eq. 76.3 in \cite{Landau})
\begin{equation}
{\bf f} =    \frac{2}{3}  r_0 e \gamma (\btt \nabla ) {\bf E} + \frac{2}{3} r_0^2 (\btt {\bf E}){\bf E} -
\frac{2}{3} r_0^2 \gamma^2 \btt \left[ \mbox{\rm E}^2 -(\btt {\bf E})^2 \right]  ,
\label{LL_eq}
\end{equation}
where $\btt = {\bf v}/c$, ${\bf v}$ is the velocity of an electron, $c$ is the velocity of light, $m$ is  the mass  of an electron, $r_0 = e^2/mc^2$ is a classical electron radius, $\gamma = (1-\beta^2)^{-1/2}$ is the Lorentz factor, $\mbox{\rm E}=\mid \! \! {\bf E} \! \! \mid$ , 
$\nabla = (\partial / \partial x, \partial / \partial y, \partial /\partial z )$    is the spatial gradient. 

It is convenient to rewrite Eq. (\ref{LL_eq}) in the form with a total time derivative
\begin{equation}
{\bf f} = \frac {2e}{3c}r_0 \frac{d}{dt} (\gamma  {\bf E})  + \frac{1}{c} \btt \left(  \frac{d W}{dt} \right)_{cl} ,
\label{LL_eq1}
\end{equation}
where  $(dW/dt)_{cl}= (2/3)ce^2g^2$ represents energy losses due to the radiation per unit time, $g^2$ is a square of the 4-acceleration
\begin{equation}
g^2 \equiv g_i g^i = - \frac{1}{c^2} \gamma^4 \left[  \dot{\btt}^2 + \gamma^2 (\btt \dot{\btt})^2   \right], 
\label{4_acceleration2}
\end{equation}
where $g^i=du^i/ds$, $u^i$ are components of the dimensionless 4-velocity, $u^2=1$, $i=(0,1,2,3)$,  $ds=cd\tau$ is a proper interval, $\tau$ is a proper time, $\dot{\btt} = d \btt /dt$ is a 3-acceleration divided by the velocity of light. It has been taken into account  in the above equations that $d{\bf E}/dt = c (\btt \nabla ) {\bf E}$, since the field is static, i.e. $\partial {\bf E} / \partial t = 0$, and  that acceleration is expressed in terms of the external field as  
\begin{equation}
\dot{\btt} = \frac{e}{m c \gamma} \left[  {\bf E}  - \btt  (\btt {\bf E}) \right]. 
\label{acceleration1}
\end{equation}

The self-force does the work which per unit time is equal to the product ${\bf fv}$  and  defines the reversible and irreversible emitted radiation power.  In the case considered it can be expressed in the form containing the  total time derivative
\begin{equation}
{\bf fv} = \frac {2}{3}e r_0 \frac{d}{dt} \left[ \gamma  (\btt {\bf E}) \right] +   \left(  \frac{dW}{dt} \right)_{cl} .
\label{Power_LL}
\end{equation}
The energy radiated throughout the entire trajectory is determined only by the second term, since  the external field vanishes at infinity such that the integral of the first term over time is zero. The second term being a Lorentz invariant  represents the irreversible  energy loss and  is called  the ``Li\'enard term'', whereas the term with a total time derivative is called ``the acceleration term'' (page 253 in  (\cite{Schott})), or the ``Schott term''.  By analogy the first term in Eq. (\ref{LL_eq1}) may also be called the ``Schott term''. 
This definition is justified by the fact that after multiplying Eq. (\ref{LL_eq1}) by the velocity, the extra addend that appears from the first term and contributes to the second term in Eq. (\ref{Power_LL}) is proportional to $\gamma^{-2}$, and  becomes negligible when $\gamma \gg 1$.  
The   second  term in  Eq. (\ref{LL_eq1})  is always negative, directed opposite to the electron's velocity  and  represents  the ``Li\'enard term'' in the self-force formula.     

The LL equation is an approximate equation, when the trajectory quantities  in the more accurate Lorentz-Abraham-Dirac (LAD) formula (Eq. 76.2 in \cite{Landau})  are expressed in terms of the  external electromagnetic field that is not perturbed by the field of radiation. The separation of the radiative self-force and the radiation power into the Schott and Li\'enard contributions  takes place also in the general case of the LAD equation. 
The total time  derivatives in Eqs. (\ref{LL_eq1})  and (\ref{Power_LL}) correspond to the spatial and time components of the total derivative term in LAD equation, equal to $(2/3) e^2 dg^i/ds$.

It may seem that the first term in Eq. (\ref{Power_LL}) is not observable, since for all realistic trajectories it vanishes when integrated over time.
This is true if the movement of an electron occurs along a trajectory  known a priori. The  form of this trajectory, however, depends on whether the Schott term  has been taken into account in the equations of motion with account of the radiation self-force  in Eq. (\ref{LL_eq1}). This idea is the basis of experimental verification of the classical radiation self-force effects \cite{Ulrik2017}. As it will be shown below, such effects appear in channeling of relativistic electrons in oriented crystals in the form of radiative damping of their transverse energy.

\section{Radiation self-force  in channeling }

Relativistic electrons  moving nearly parallel to the crystallographic axis  or planes  generate strong electromagnetic radiation in the X-ray or gamma  range \cite{Beloshitsky1982}, \cite{Ulrik2005}. The motion of an electron is governed by the electrostatic potential depending only on the distance to the axis (plane) $U({\bf r}_\perp)$, where in the axial case  
${\bf r}_\perp = (x,y)$ is two-dimensional transverse coordinate perpendicular to the atomic axis. In the case of planar channeling it is one-dimensional, such that  $U=U(x)$, where $x$ is the distance to the atomic plane. The longitudinal coordinate $z$ coincides with the atomic  axis (plane).  Typical values of the axis potential are $U \sim 2Ze^2 /d \approx 10^2$ eV.
The electric field strength acting on the electron is $e{\bf E}= - \nabla_\perp U$, $\nabla_\perp = (\partial / \partial x, \partial / \partial y)$.  The longitudinal component of this field is absent ${\rm E}_z =0$.  

Classical radiation power emitted by the channeled electron is
\begin{equation}
\left(  \frac{dW}{dt} \right)_{cl} = - \frac{2c}{3e^2} r_0^2 \gamma^2 \mid \nabla_\perp U \mid^2 .
\label{class_chan_rad_power}
\end{equation}
The transverse and longitudinal components of the self-force in  Eq.(\ref{LL_eq}) for channeled electrons  are
\begin{eqnarray}
{\bf f}_\perp &=&    \frac{2}{3}  r_0 e \gamma (\btt_\perp \nabla_\perp ) {\bf E} + 
 \frac{2}{3} r_0^2 (\btt_\perp {\bf E}){\bf E} +   \label{LL_eq_chann1}  \\ 
&+&  \btt_\perp \left(  \frac{dW}{dt} \right)_{cl} +  
\frac{2}{3} r_0^2 \gamma^2 \btt_\perp   (\btt_\perp {\bf E})^2   , \nonumber \\
f_z &= &\beta_z  \left(  \frac{dW}{dt} \right)_{cl}.
\label{LL_eq_chann2}
\end{eqnarray}

Let us estimate the relative contribution of four terms in Eq.(\ref{LL_eq_chann1}) for ultrarelativistic motion $\gamma \gg 1$. The force acting on an electron is  $\mid \! \nabla_\perp U \! \mid \approx U_0/a_F$, 
$a_F$ is the Thomas-Fermi screening parameter, $a_F \approx$(0.1-0.2) \AA, $U_0$ is the potential depth. The transverse velocity $\beta_\perp \ll 1$ is equal to the angle of deflection of an electron by the atomic string $\theta_e$. It is  of the same order of magnitude as the critical channeling angle $\theta_L$, i.e.  $\beta_\perp \sim \theta_L$, $ \theta_L = (4Ze^2 /d E )^{-1/2}$, $ E = \gamma mc^2$ is electron's energy (not to be confused with the notation for the electric field {\rm E}).  We see, therefore, that dominating are the first and the third terms in Eq. (\ref{LL_eq_chann1}), which define the relative contribution of the  Schott and Li\`enard terms  
\begin{equation}
\eta \equiv \frac{\mbox{Li\`enard term}}{\mbox{Schott term} } = \gamma \frac{U_0}{mc^2}. 
\label{L_to_S}
\end{equation}
The second and the last terms  in Eq. (\ref{LL_eq_chann1}) are $\gamma^2$  and  $\theta_L^{-2}$ times less than the third one, correspondingly.  In the case of axial channeling $\eta \sim 1$ at energies of 10--30 GeV. At smaller energies, when the dipole approximation is valid, i.e. $\theta_L \gamma \ll 1$,  the first term (Schott term) in Eq.(\ref{LL_eq_chann1}) dominates, whereas at higher energies dominating is the Li\`enard term. The longitudinal component of the velocity for arbitrary ultrarelativistic motion  is expressed through the transverse velocity  $\beta_z (t) \approx 1 - \gamma^{-2}/2 -\btt_\perp^2 (t)/2$, where the terms $\sim \beta^4_\perp$ and $\gamma^{-4}$ are neglected. 

The equations of motion of channeled electrons  can be written as 
\begin{eqnarray}
\frac{d{\bf p}_\perp}{dt}  &= &  - \nabla_\perp U + {\bf f}_\perp   \label{LL_eq_chann1a}  \\ 
\frac{d p_z}{dt}  &= &  \frac{\beta_z}{c}  \left(  \frac{dW}{dt} \right)_{cl}, 
\label{LL_eq_chann2a}
\end{eqnarray}
where ${\bf p} = \gamma mc \btt$ is electron momentum and
\begin{equation}
{\bf f}_\perp =    \frac{2}{3}  r_0 e \gamma (\btt_\perp \nabla_\perp ) {\bf E} + 
  \btt_\perp \left(  \frac{dW}{dt} \right)_{cl} .
\label{LL_eq_chann}
\end{equation}
The first term in this formula agrees with Eq. (3) in \cite{Ulrik2017} written for planar channeling.

All expressions given above are valid for arbitrary ultrarelativistic motion, $\gamma \gg 1$, with additional condition that electron's energy must sufficiently exceed the potential of its interaction with an external field $E \gg \mid \!\! U \!\! \mid$. This case is usually of practical interest. The channeling angle $\theta_L$ in above formulas should be replaced by the angle of deflection of an electron  by the external field $\theta_e$, $\theta_L \rightarrow \theta_e$, such that the transverse velocity always remains non-relativistic $\beta_\perp \approx \theta_e \ll 1 $.

\section{Transverse energy damping in channeling}
In the absence of the radiation and  multiple scattering the transverse energy of an electron becomes an integral of the transverse motion, since the continuum atomic string (plane) potential does not depend on the longitudinal coordinate. Multiple scattering is of stochastic nature and   will not be considered in what follows. The transverse energy $\varepsilon $ of an electron is
\begin{equation}
\varepsilon = \frac{{\bf p}_\perp^2}{2m\gamma} + U({\bf r}_\perp) .
\label{tr_energy}
\end{equation}
The radiation leads to the transverse energy change with time. The transverse momentum is a rapid function of time  and  varies with the frequency of transverse oscillations, while the total energy varies with time adiabatically due to the radiation.  The total time derivative of Eq. (\ref{tr_energy}) with account of the equations of motion (\ref{LL_eq_chann1a}) and Eq. (\ref{LL_eq_chann}) give
\begin{equation}
 \frac{d \varepsilon}{dt} = \frac{2}{3}  r_0 e \gamma c \btt_\perp (\btt_\perp \nabla_\perp ) {\bf E} + 
 \frac{1}{2}  \btt_\perp^2  \left(  \frac{dW}{dt} \right)_{cl} .
\label{tr_energy_damping}
\end{equation}
The second ``Li\'enard'' term in this equation is always negative and dominates at high energies when $\eta \gg 1$.    The first ``Schott''  term can be both negative and positive. Further it will be clear  that the first term in Eq.(\ref{tr_energy_damping}) becomes also always negative being averaged over the transverse motion period. 


Consider the simplest case of one dimensional transverse motion during the planar channeling. In this case  Eq. (\ref{tr_energy_damping}) gives 
\begin{equation}   
 \frac{d \varepsilon}{dt} = \frac{2}{3}  r_0  \gamma c \beta_x^2 U''  + 
 \frac{1}{2}  \beta_x^2  \left(  \frac{dW}{dt} \right)_{cl},
\label{Planar_damping}
\end{equation}
where $U'' (x) \equiv d^2U/dx^2$, and $x$ is a transverse coordinate. 

 Eq. (\ref{Planar_damping}) should be  averaged over the equilibrium distribution of channeled particles with transverse energy $\varepsilon$    over the transverse coordinate distribution function  $dw(x) = 2dx/(v_x T)$, which coincides with averaging over the period of the  transverse motion  $T$   
\begin{equation}
\left\langle ... \right\rangle \equiv \frac{2}{T} \int_{x_{min}}^{x_{max}} (...) \frac {dx}{v_x} =
 \frac{\sqrt{2E}}{cT} \int_{x_{min}}^{x_{max}} (...) \frac {dx}{\sqrt{\varepsilon - U(x)}},
\label{planar_averaging}
\end{equation}
with 
\begin{equation}
T =  \frac{\sqrt{2E}}{c} \int_{x_{min}}^{x_{max}}  \frac {dx}{\sqrt{\varepsilon - U(x)}},
\label{planar_period}
\end{equation}
where $x_{min}$ and $x_{max}$ are solutions of the equation $\varepsilon = U(x)$  (turning points). 

Integration by parts gives for the average value of $\beta_x^2 U''$  in Eq. (\ref{Planar_damping})
\begin{equation}
\left\langle    \beta_x^2 U''  \right\rangle = 
- \frac{2}{cT} \int_{x_{min}}^{x_{max}}  U' \dot{\beta}_x  \frac {dx}{v_x }.
\label{aver_pl1}
\end{equation}
It has been taken into account here that the transverse velocity vanishes in the turning points $v_x (x_{min})= v_x (x_{max})=0$. With a good approximation we can assume also  that according to the equations of the transverse motion without radiation 
$\dot{v}_x =-  U' / (m\gamma)$. This gives $\left\langle    \beta_x^2 U''  \right\rangle = \left\langle ( U')^2  \right\rangle /E$.   

The Schott term contribution  in Eq. (\ref{Planar_damping}) averaged over the transverse motion period becomes 
\begin{equation}
\left\langle    \frac{2}{3}  r_0  \gamma c  \beta_x^2 U''  \right\rangle = 
\frac{1}{\gamma^2}  \left\langle     \left(  \frac{dW}{dt} \right)_{cl}    \right\rangle .
\label{Schott_term_planar_damping}
\end{equation}
This term is always negative.

Finally for the mean value of the transverse energy change due to the  channeling radiation we obtain 
\begin{equation}
\left\langle    \frac{d \varepsilon}{dt}   \right\rangle = 
\frac{1}{\gamma^2}  \left\langle     \left(  \frac{dW}{dt} \right)_{cl}    \right\rangle + 
\left\langle    \frac{( \varepsilon - U)}{E}   \left(  \frac{dW}{dt} \right)_{cl}  \right\rangle ,
\label{averaged_rad_damping}
\end{equation}
where we have taken into account that $\beta_x^2 (x) = 2[\varepsilon - U(x)]/E$. Eq. (\ref{averaged_rad_damping}) for planar channeling  was derived at the dawn of the channeling radiation physics \cite{Bazylev1979}.

Axial channeling of electrons can be considered as a motion in axially symmetric field of the atomic string with the potential $U(r)$, where $r$ is the transverse distance to the string $r=\mid \!\! {\bf r}_\perp \!\! \mid$.  In this case one more integral of the transverse motion exists. It is the angular momentum of an electron with respect to the string $\mu = xp_y -yp_x = m\gamma rv_\varphi$, $\varphi$ is the  azimuth angle of an electron in the transverse plane, $v_\varphi = rd\varphi /dt = \mu c^2 /(Er)$ is the corresponding component of the transverse velocity, the radial part of which is $v_r=dr/dt$, $v_\perp^2 = v_r^2 +v_\varphi^2$.   
 It can be shown that Eq. (\ref{averaged_rad_damping}) is valid in this case too, where the averaging is over the period of the radial transverse oscillations 
\begin{equation}
\left\langle ... \right\rangle \equiv 
 \frac{\sqrt{2E}}{cT} \int_{r_{min}}^{r_{max}} (...) \frac {dr}{\sqrt{\varepsilon - U_{eff}(r)}},
\label{axial_averaging}
\end{equation}
where $U_{eff}(r) = U(r) + \mu^2c^2/2Er^2$ is an effective potential; $ r_{min}$  and $r_{max}$ are radial turning points defined by the equation $\varepsilon = U_{eff} (r)$. 

 For the angular momentum damping in axial electron channeling one can obtain 
\begin{equation}
\left\langle    \frac{d \mu}{dt}   \right\rangle =  \frac{\mu}{E} 
\left[  - \frac{2}{3} r_0 c \gamma  \left\langle    \frac{U'}{r}   \right\rangle   +  
 \left\langle     \left(  \frac{dW}{dt} \right)_{cl}    \right\rangle  \right],
\label{ang_momentum_damping}
\end{equation}
where the first and second terms  in the square brackets  represent  Schott and Li\'enadr contributions.  The latter one is always negative, while the first one can be both -- positive and negative.  

We have considered channeled electrons whose transverse trajectories  are finite and transverse energy is negative $\varepsilon <0$. Eq. (\ref{averaged_rad_damping})  is valid for quasi-channeled electrons as well ($\varepsilon >0$), where the angle brackets for planar channeling does mean the averaging according to the Eq. (\ref{planar_averaging}) with $x_{min}$, $x_{max}$ $=\pm d_p/2$, $d_p$ is the distance between the atomic planes. For axially quasi-channeled electrons we must average over the whole transverse trajectory around a single atomic string, where $T$ becomes the time of the interaction.     

In real crystals, electrons experience strong multiple scattering by thermal vibrations of atoms \cite{mkh1982}, \cite{Beloshitsky1986_Phys_Rep}. This leads to a stochastic change in the transverse energy and angular momentum.  The corresponding terms should be added to the right hand side of the Eqs. (\ref{averaged_rad_damping}) and (\ref{ang_momentum_damping}). In what follows we shall disregard this factor. 

\section{Correspondence to the quantum theory}

At electron energies of several MeV, their transverse wavelength, $\lambda_\perp \approx 2\pi \hbar c/E\theta_L$,
is comparable with the distance between atomic planes (axes), which leads to the appearance of several quantum transverse energy levels  \cite{jua1983}. The number of levels increases with an increase in the electron energy, so that at energies above several tens of MeV, the transitions between individual levels are no longer distinguishable and the emission spectrum is described by classical electrodynamics. The formulas for radiation probabilities are especially simple in the dipole approximation when the angle of the deflection of an electron by the external field  is much less than the characteristic radiation angle $\theta_\gamma \approx \gamma^{-1}$, i.e. when $\theta_L \gamma \ll 1$ (or $\eta \ll 1$).  

The frequency distribution of the probability of the channeling radiation due to the transition from the state with transverse energy $\varepsilon_i$ to the state with final transverse energy  $\varepsilon_f$ is (see, for example, Eq. (123) in \cite{Beloshitsky1982})  
\begin{equation}
\frac{dw_{fi}}{d\omega} = \frac{\alpha}{c^2} \mid ({\bf r}_\perp)_{fi}\mid^2 \Omega_{fi}^2 F \left(  \frac{\omega}{\omega_m} \right), \, \, \omega \le \omega_m, 
\label{dipole_rad_probability}
\end{equation}
where $\alpha = 1/137$ is a fine structure constant, 
$({\bf r}_\perp)_{fi} \!= \! \langle \varepsilon_f \! \!  \mid \!\! {\bf r}_\perp \!\!   \mid \!\! \varepsilon_i \rangle$ is a matrix element of the transverse coordinate, $\Omega_{fi} = (\varepsilon_i - \varepsilon_f)/\hbar$, $\omega_m = 2 \gamma^2 \Omega_{fi}$,  $F(x)=1-2x+2x^2$ is a dipole radiation frequency factor, $x \le 1$.  

The rate of the transverse energy change due to the photon emission is
\begin{equation}
\frac {d \varepsilon}{dt} = \sum_{f<i} \int \hbar \Omega_{fi} \frac {dw_{fi}}{d\omega} d\omega.
\label{tr_energy_change_due_to_rad}
\end{equation}
Making use of Eq. (\ref{dipole_rad_probability}) we obtain
\begin{equation}
\frac {d \varepsilon}{dt} = \frac {4e^2}{3c^3}  \gamma^2  \sum_{f<i}  \Omega_{fi}^4  \mid ({\bf r}_\perp)_{fi}\mid^2   .
\label{tr_energy_change_due_to_rad1}
\end{equation}
In the same way we obtain the rate of the total energy change due to the dipole radiation
\begin{eqnarray}
\frac {d E}{dt} & = & \sum_{f<i} \int \hbar \omega  \frac {dw_{fi}}{d\omega} d\omega  \nonumber \\
& = & \frac {4e^2}{3c^3}  \gamma^4  \sum_{f<i}  \Omega_{fi}^4  \mid ({\bf r}_\perp)_{fi}\mid^2  .
\label{total_energy_change_due_to_rad}
\end{eqnarray}
Comparison of Eqs. (\ref{tr_energy_change_due_to_rad1}) and (\ref{total_energy_change_due_to_rad}) leads to the result
\begin{equation}
\frac {d \varepsilon}{dt} = \frac {1}{\gamma^2} \frac {d E }{dt}.
\label{tr_total_energy_change}
\end{equation}

The quantum dipole result Eq. (\ref{tr_total_energy_change}) exactly coincides with that given by the classical Schott term in Eq. (\ref{averaged_rad_damping}). 
Consequently, we can conclude that the Schott term  is nothing but the classical analogue of the phenomena  associated with quantum transitions between transverse energy levels. The same consideration is valid for quasi-channeled electrons with infinite transverse classical trajectories. In this case quantum transitions occur between the states of a continuous spectrum.

Consider a more general case beyond the dipole approximation. The energy and longitudinal momentum  conservation laws due to the photon emission are: $E_f = E_i - \hbar \omega$, $p_{z,i} = p_{z, f} + \hbar k \cos \theta$; where $k=\omega /c$, $\theta$ is the angle between $z$-axis and the direction of the photon emission. These equations give the following result for the transverse energy change due to the photon emission    
\begin{equation}
\hbar \Omega_{fi} \equiv  \Delta \varepsilon (\omega) = \frac {\hbar \omega}{2\gamma^2} \frac {E}{E-\hbar \omega} +
\hbar \omega \frac {\theta^2}{2},   
\label{tr_en_change_conservation_laws}
\end{equation}
where terms $\sim \theta^4$ and $\gamma^{-4}$ have been neglected. Eq. (\ref{tr_en_change_conservation_laws}) has been derived for axial case. All subsequent conclusions are valid for the planar case as well.

Eq. (\ref{tr_en_change_conservation_laws}) written in  different  forms was obtained earlier \cite{mkh1993} (see also  Eq. (99) in \cite{Beloshitsky1982}). If $w(\omega)d\omega$ is a differential probability of photon emission per unit time, than the rates of the transverse and total energy change are
\begin{eqnarray}
\frac {d \varepsilon}{dt} & = & - \int \Delta \varepsilon (\omega) w(\omega) d\omega ,
\label{general_tr_energy_change} \\
\frac {d E }{dt} & = & - \int  \hbar  \omega  \, w(\omega) d\omega .
\label{general_total_energy_change} 
\end{eqnarray}
In the case opposite to the dipole approximation, when $\theta_l \gamma \gg 1$, the probability $w(\omega)$ is defined by the well known quantum synchrotron-like constant field approximation formula  \cite{Baier1998}. For non-uniform fields $w(\omega)$  is given in \cite{mkh_Nitta_2002}.  Since $\Delta \varepsilon$  can depend also on angles of the photon emission,  the integration in Eq. (\ref{general_tr_energy_change}) should be preformed over the angular variables as well. 

For relatively small energies when the dipole approximation is valid,  i.e. $\theta_l \gamma \ll 1$, and  emitted photons are soft, $\hbar \omega \ll E$,  the first term in Eq.(\ref{tr_en_change_conservation_laws}) is approximately equal to $\hbar \omega /(2\gamma^2)$. In the second term we can replace $\theta \approx 1/\gamma$. 
We therefore obtain formula (\ref{tr_total_energy_change}) from Eqs.(\ref{general_tr_energy_change}) and (\ref{general_total_energy_change}). In this limit both terms in Eq. (\ref{tr_en_change_conservation_laws}) give equal contribution to the Schott-like term. In the opposite limit when the angle $\theta_e$  of the deflection of an electron by the external field is much larger than the angle of photon emission, i.e.  $\theta_e \gg 1/\gamma$ ($\eta \gg 1$), we can disregard the first term in Eq.(\ref{tr_en_change_conservation_laws}) (for photon energies not very close to the energy of the electron) and replace,  $\theta \approx \theta_e = \beta_\perp$,  in the second term. I.e. in this limit we assume that photon is emitted exactly in the direction of electron motion. This gives the transverse energy change due to the emission of the photon with the energy $\hbar \omega$
\begin{equation}
\Delta \varepsilon (\omega) = - \frac {\varepsilon - U}{E} \hbar \omega.     
\label{delta_epsilon_omega}
\end{equation}
After the averaging over the radiation probability,  Eq.(\ref{delta_epsilon_omega}) becomes in agreement with the second term in formula (\ref{averaged_rad_damping}). We see, that in the limit, $\theta_L \gamma \gg 1$, the second term in Eq. (\ref{tr_en_change_conservation_laws}) corresponds to the Li\'enard-like contribution to the radiation damping. In the computer simulations of channeling radiation of above 100 GeV electrons only the term Eq.(\ref{delta_epsilon_omega}) was usually taken into account \cite{mkh1993}, \cite{mkh1996}.  
 
A special case happens for hard emitted photons when $\hbar \omega \rightarrow E$. According to the Eq. (\ref{delta_epsilon_omega})  the transverse energy change in this case becomes infinite due to the first term. In reality, however, 
the probability of photon emission $w(\omega)$ falls down exponentially when $\hbar \omega \rightarrow E$, and the integral in Eq. (\ref{general_tr_energy_change}) remains finite. The effect of the first term in Eq. (\ref{delta_epsilon_omega}) is not significant  for electron energies achieved in laboratory at present time.  For above TeV energies, however, this factor can be important  because at such energies the probability of the spin-flip radiation becomes quite strong for photons with  $\hbar \omega \sim E$ \cite{Ulrik_spin_flip}, \cite{mkh2006}.

The present analysis of the Eq. (\ref{tr_en_change_conservation_laws}) permits one to generalize the classical Eqs. (\ref{averaged_rad_damping}) and (\ref{ang_momentum_damping}).  For example, to take into account the quantum recoil  due to the  hard photon emission and influence of spin we can replace $(dW/dt)_{cl} \rightarrow (dW/dt)_{quant}$, where $(dW/dt)_{quant}$ is defined by the Eq. (\ref{general_total_energy_change}) with some general expression for radiation probability $w$, given in \cite{Ulrik_spin_flip} and \cite{mkh2006}. 

\section{Summary and concluding remarks}  

If the entire electron trajectory is completely defined, then the total energy of electromagnetic radiation from this trajectory is determined exclusively by the second term in Eq. (\ref{Power_LL}), that is, the Li\'enard contribution. The role of the radiation self-force is that it affects the trajectory. Taking this force into account in the equations of motion leads to a difference in the trajectory from that which would be without taking this force into consideration, and therefore the emission spectrum will also differ. At relatively low energies, when $\eta \ll 1$, the influence of the radiative self-force on the trajectory is completely due to the first term in Eq. (\ref{LL_eq1}), that is, the Schott term. In the opposite limit, when $\eta \gg 1$, everything is determined by the second term in Eq. (\ref{LL_eq1}) (Li\'enard  term). The condition $\eta \ll 1$  coincides with the condition of dipole radiation, when the angle of electron deflection by the external field is much less than the characteristic angle of radiation.

It is convenient to study the effects of radiation on the motion for ultrarelativistic electrons, $\gamma \gg 1$, since in this case the radiation  is quite intense, significantly affects the motion and is measurable with high accuracy. Channeling radiation in oriented crystals is unique in that sense that it has a pronounced quantum character in two extreme cases of low and high energies.

At low energies of few MeV ($\eta \ll 1$)  channeling radiation is due to the dipole quantum transitions with well-pronounced lines in the emission spectrum. 
These transitions are a quantum analog of the classical Schott term, i.e. the  first term with the total time derivative in Eq. (\ref{LL_eq1}). 
 Note that it is wrong to say that the radiation is due to the Schott term. Schott term affects the trajectory and, through this, the emission spectrum. At electron energies of several MeV, the change in the transverse energy due to the quantum transitions is a significant fraction of the  transverse potential barrier. This fraction decreases with increasing the energy of an electron when the role of the second term in Eq. (\ref{LL_eq1}) (Li\'enard  term)  becomes dominating. This term describes a decrease in transverse energy, approximately proportional to a decrease in total energy, i.e. $\Delta \varepsilon / \varepsilon  \simeq  \Delta E /E   $. At low energies, the role of this factor is negligible, but it becomes decisive when $\eta \gg 1$. In this limit, at electron energies above 100 GeV, the quantum effects of photon recoil and influence of electron spin on the radiation probability begin to dominate.

 The author is  grateful to J.U. Andersen and U.I. Uggerh\o j  for useful discussions and interest in this work.


\end{document}